\documentstyle [12pt] {article}
\newcommand{\cs}[3]{{{#3} \brace {#1 #2}}}

\newcommand{\h}[1]{\mathop{\lambda}\limits_{#1}\ \!\!\!}
\newcommand{\C}[1]{\mathop{C}\limits_{#1}\ \!\!\!}

\newcommand{\edf}{\ {\mathop{=}\limits^{\rm def}}\ }

\begin{document}
\begin{center}
\bf {AN AP-STRUCTURE WITH FINSLERIAN FLAVOR: I }\\

\end{center}
\begin{center}
\bf{M.I.Wanas\footnote{Astronomy Department, Faculty of Science,
Cairo University, Giza, Egypt.

E-mail:wanas@frcu.eun.eg}}
\end{center}

\begin{abstract}
A geometric structure (FAP-structure), having both absolute
parallelism and Finsler properties, is constructed. The building
blocks of this structures are assumed to be functions of position
and direction. A non-linear connection emerges naturally and is
defined in terms of the building blocks of the structure. Two linear
connections, one of Berwald type and the other of the Cartan type,
are defined using the non-linear connection of the FAP. Both linear
connections are non-symmetric and consequently admit torsion. A
metric tensor is defined in terms of the building blocks of the
structure. The condition for this metric to be a Finslerian one is
obtained. Also, the condition for an FAP-space to be an AP-one is
given.

\end{abstract}
\section{Motivation}
It is well known that the General Theory of Relativity (GR), the
best known theory describing gravitational interactions, is
suffering from many problems nowadays. For example, it cannot
account for the flatness of the rotation curves of spiral galaxies
[1]. Also, there are some problems concerning the interpretation of
the accelerated expansion of the Universe [2] discovered using SN
type Ia observations [3]. Since GR is successful in the domains of
the solar system and binary pulsars, which have relatively small
scales, some authors believe that its problems are scale problems
(cf.[2]). Consequently, a different theory is required in order to
deal with diverse ranges of scale.

 The origin of problems of GR can be seen
from a different point of view. The field equations of GR can be
written as (cf.[4])
$$
R_{\mu \nu}- \frac{1}{2}g_{\mu \nu} = - \kappa T_{\mu \nu},
\eqno{(1)}
$$
 where the L.H.S. is Einstein's tensor derived in the context of
Riemannian  geometry while the R.H.S. is a phenomenological tensor
scaled by the constant $\kappa$. The later tensor is not a member of
the geometric structure used. It describes the material-energy
distribution in the system under consideration. In the case of empty
space i.e. in the absence of material distribution, the field
equations (1) will reduce to
$$
R_{\mu \nu} = 0. \eqno{(2)}
$$
It is to be considered that all the success of GR comes from the use
of (2), not (1), in application.  Many of the problems of GR are
connected to  the use of (1) in applications. Einstein was the first
who directed the attention to solve this problem [5]. This is
manifested in his series of attempts to construct what has been
called a "{\it{Unified Field Theory}}", in which tensors describing
the material-energy distribution and fields are members of the
geometric structure used. Some authors have attempted to follow his
scheme for solving the problems of GR (cf. [6], [7]). The attempts
in this direction concentrate on using different geometries of more
wider structures. This is because the ten field variables $(g_{\mu
\nu})$ of the 4-dimensional Riemannian geometry are just sufficient
to describe
 gravity alone. So, in order to accommodate more physical
quantities and interactions, some authors have used extra dimensions
({\it{e.g. Kaluza-Klein type Theories)}}. Others, have used
different geometries ({\it{e.g. Riemann-Cartan, Wyel-Cartan,
Absolute Parallelism}}). Absolute Parallelism (AP) geometry is more
wider than the Riemannian one in the sense that, in 4-dimensions, it
has six more degrees of freedom. Another advantage of the use of the
AP-space {\footnote{This geometric structure is known as
"Parallelisable Manifold", relative to geometers.}}, in application,
is that it has an associated Riemannian space. This facilitates  a
direct comparison of any theory, written in the AP-space, with GR.
On the other hand, attention  has been recently directed to Finsler
geometry and its generalization to gain more insight on the
infrastructure of physical phenomena (cf. [8], [9]).

The aim of the present work is to construct an AP-structure with
Finslerian properties,
 in particular, an AP-structure
with an associated
Finsler space.  This structure would facilitate:\\
{1.} The use of the advantages of both machineries of AP and Finsler
geometries. \\
{2.} The gain of more information on the infrastructure of physical phenomena studied in

AP-geometry. \\
{3.} Attribution of some physical meaning to objects of Finsler
geometry, since AP-geometry

has been used in many successful applications (for a
brief review cf. [10]). \\
{4.} Solution of some of the problems of GR , mentioned above, if
possible. \\ The paper is arranged as follows: In subsection 2.1 we
give a brief review of some basic formulae of the AP-structure. In
subsection 2.2. we give necessary formulae for Finsler structure. In
Section 3 we construct an AP-structure using the machinery of
Finsler geometry. Some concluding remarks are given in section 4.
\section{Basic Formulae}
In this section, we are going to give a brief review on basic
formulae necessary for comparison with the results obtained in
section 3.
\subsection{Absolute Parallelism Space}
An AP-space $(M,\lambda)$ is an n-dimensional differentiable
manifold $(M)$ equipped with a set of n-linearly independent vectors
$\h{i}_{\mu}$ (in what follows we are going to use Latin indices for
vector numbers and Greek indices to indicate coordinate components.
Each type of indices runs from 1 to $n$ ). Since these vectors are
linearly independent, the determinant of matrix $(\h{i}_{\mu})$ is
non-vanishing. Consequently, we can define the contravariant
components of $\h{i}_{\mu}$ such that {\footnote{Latin indices
(vector numbers) are always written in a lower position neither
covariant nor

contravariant. Summation convention is applied over repeated indices
whatever their positions.}} ( for more details cf.[11]) $$
 \h{i}_{\alpha}\h{i}^{\beta} =
\delta^{\beta}_{\alpha},\eqno{(3)}
$$
$$
\h{i}_{\alpha}\h{j}^{\alpha} = \delta_{ij}. \eqno{(4)}
$$
A linear connection is defined as
$$
\Gamma^{\alpha}_{.\mu \nu} \edf \h{i}^{\alpha} \h{i}_{\mu ,
\nu},\eqno{(5)}
$$
where the comma (,) is used here to characterize ordinary partial
differentiation relative to $x^{\nu}$. It is clear that the
connection (5) is non-symmetric. So, one can define a torsion as
$$
\Lambda^{\alpha}_{.\mu \nu} \edf \Gamma^{\alpha}_{.\mu \nu}-
\Gamma^{\alpha}_{.\nu \mu}, \eqno{(6)}
$$
which is a tensor of type $(1,2)$, skew symmetric in its two lower
indices. Now, using the vectors $\h{i}$ , the following second order
symmetric tensors can be defined,
$$
g_{\mu \nu}  \edf \h{i}_{\mu}\h{i}_{\nu},  \eqno{(7)}
$$
$$
g^{\alpha \beta}  \edf \h{i}^{\alpha}\h{i}^{\beta}. \eqno{(8)}
$$
It is clear that these tensor, using (3), (4), satisfy the relation,
$$
g_{\mu \alpha} g^{\mu \beta} = \delta^\beta_\alpha.
$$
Using the above properties and knowing that $g_{\mu \nu}$ is non
degenerate, then $(7)$ can be used as a metric tensor of a
Riemannian space associated with any AP-structure. Also, the tensors
(7) and (8) can be used to lower or raise tensor indices. Using (7)
and (8) we can define as usual a linear symmetric connection, which
is Christoffel symbol of the second kind  $$ \cs{\mu}{\nu}{\alpha}
\edf \frac{1}{2}g^{\alpha \beta} (g_{\beta\mu, \nu}+ g_{\beta\nu,
\mu}-g_{\mu \nu, \beta}). \eqno{(9)}
$$
We can use the connections (5) and (9) to perform covariant
differentiation as follows,
$$
A_{\stackrel{\mu}{+}| \nu} \edf A_{\mu, \nu} -A_{\alpha}
\Gamma^{\alpha}_{.\mu \nu},\eqno{(10)}
$$
$$
A_{\mu; \nu} = A_{\mu, \nu} - A_{\alpha} \cs{\nu}{\mu}{\alpha},
\eqno{(11)}
$$
where $A_{\mu}$ is an arbitrary covariant vector.

Applying the types of differentiation given by (10) and (11) to the
vectors $\h{i}$ and the tensor (7) we get the following results,
$$
\h{i}_{\stackrel{\mu}{+} | \nu} =0, \eqno{(12)}
$$

$$
g_{\mu \nu | \stackrel{\sigma}{+}} =0, \eqno{(13)}
$$

$$
g_{\mu \nu ; \sigma} = 0. \eqno{(14)}
$$

Equations (12) indicates absolute parallelism while (13) and (14)
are metricity conditions. The tensors $g^{\mu \nu}$ and $g_{ \mu
\nu}$ can be used to raise and lower coordinate indices under the
above covariant differentiation signs.

A third order tensor ,{\it {contortion}}, can be defined as
$$
\gamma^{\alpha}_{. \mu \nu} \edf \h{i}^{\alpha}\h{i}_{\mu;
\nu},\eqno{(15)}
$$
which is non-symmetric with respect to its lower two indices. It can
be shown that
$$
\Lambda^{\alpha}_{. \mu \nu}=\gamma^{\alpha}_{. \mu \nu}-
\gamma^{\alpha}_{. \nu \mu} , \eqno{(16)}
$$
$$
C_{\mu}=\Lambda^{\alpha}_{. \mu \alpha}= \gamma^{\alpha}_{. \mu
\alpha} , \eqno{(17)}
$$
where $C_{\mu}$ is a covariant vector, {\it the basic vector}. The
curvature tensor corresponding to (5) vanishes identically, due to
(12), but the curvature tensors corresponding to its dual
$\tilde\Gamma^{\alpha}_{\mu \nu}(= \Gamma^{\alpha}_{\nu \mu}) $ and
that corresponding to its symmetric part $\Gamma^{\alpha}_{(\mu
\nu)}$ do not vanish [11].
\subsection{Finsler Structure}
A Finsler space $(M,F)$ is an n-dimensional differentiable manifold
$M$ equipped with a scalar $F(x,y)$ function of $x(t), y(t)(=\frac
{dx}{dt})$ where $t$ is an invariant parameter (for more details
cf.[12]), $(x,y)$ are the coordinates on the tangent bundle TM. Now,
the scalar
$F(x,y)$ is assumed to satisfy the following properties: \\
1{-} $F(x,y)$ is $C^{\infty}$ on $\tau M (=TM\backslash \{0 \})$  . $F$ is called the fundamental function. \\
2{-} The function $F(x,y)$ is positively homogenous of degree one in
$y$, abbreviated as

$(P-h(1))$.\\
3{-} $y ~ \in ~ \tau M$ and transforms as,
$$
\bar{y}^\alpha = \frac{\partial \bar{x^{\alpha}}}{\partial
x^{\beta}}y^{\beta}. \eqno{(18)}
 $$
4{-} The tensor whose component defined by,
$$
g_{\alpha \beta} \edf \frac{\partial^{2} E}{\partial y^{\alpha}
\partial y^{\beta}}= E_{~ :~ \alpha \beta}, \eqno{(19)}
$$
is positive definite (non-degenerate).This tensor defines the metric
of Finsler space, where $E$ is the energy of this space and the
colon (:) is used to characterize differentiation with respect to
$y$. This tensor  is symmetric  and is $P-h(0)$. The function E is
defined by,
$$
E \edf \frac{1}{2} F^{2}. \eqno{(20)}
$$
Consequently,
$$
F^{2} \edf g_{\alpha \beta}y^{\alpha}y^{\beta}. \eqno{(21)}
$$
A tensor $C_{\alpha \beta \gamma}$  defined by,
$$
C_{\alpha \beta \gamma} \edf g_{\alpha \beta : \gamma } = E_{:
\alpha \beta \gamma}, \eqno{(22)}
$$
has the following properties: \\
{1.} It is symmetric with respect to all indices. \\
{2.} Tensor of type (0,3). \\
{3.} $P-h(-1)$. \\
  Using Euler's Theorem we get, \\
$$
C_{\alpha \beta \gamma} y^{\alpha} = C_{\alpha \beta \gamma}
y^{\beta} = C_{\alpha \beta \gamma} y^{\gamma} = 0. \eqno{(23)}
$$
Since the metric tensor is assumed non-degenerate its conjugate can
be defined such that,
$$
g^{\alpha \beta}g_{\alpha \gamma} = \delta^{\beta}_{\gamma}.
\eqno{(24)}
$$
Consequently, these tensors can be used to perform the operations of
raising and lowering indices. The following theorem can be easily
proved. \\ \\
{\bf{Theorem}:} \\ A necessary and sufficient condition for a
Finsler space to be a Riemannian  one is that the
tensor $C_{\alpha \beta \gamma}$ vanishes. \\ \\

{\bf{A Non-Linear Connection}} \\
Using the metric tensor and its conjugate, one can define the
object,
$$
G^{\alpha}_{. \beta \gamma} \edf \frac{1}{2} g^{\alpha
\sigma}(g_{\beta \sigma, \gamma}+ g_{\sigma \gamma, \beta}- g_{\beta
\gamma , \sigma}). \eqno{(25)}
$$
Since $g_{\mu \nu}$ is a function of $(x,y)$, then $G^{
\alpha}_{.\beta \gamma}$ is neither a tensor nor a connection. It is
$P-h(0)$. Consider the quantity,
$$
G^{\alpha} \edf G^{\alpha}_{. \beta \gamma}(x,y)y^{\beta}y^{\gamma}.
\eqno{(26)}
$$
 It can be shown that (26) defines a contravariant vector, called
{\bf{spray}}, and it is P-h(2). The quantity
$$
G^{\alpha}_{.\sigma} \edf G^{\alpha}_{~ :~ \sigma} = \frac{\partial
G^{\alpha}}{\partial y^{\sigma} },\eqno{(27)}
$$
 can be shown to transform according to the law,
$$
\bar G^{\alpha}_{. \beta} = \frac{\partial \bar{x}^{\alpha} }{
\partial x^{\mu}  }\frac{\partial x^{\nu}}{\partial \bar{x}^{\beta}} G^{\mu}_{ . \nu}
+\frac{\partial \bar{x}^{\alpha}}{\partial
x^{\sigma}}\frac{\partial^{2} x^{\sigma}}{\partial \bar{x}^{\beta}
\partial \bar{x}^{\gamma}}\bar{y}^{\gamma}. \eqno{(28)}
$$
So, this quantity represents the components of a non-linear
connection. It is $P-h(1)$. It can be used to define the
differential operator.
$$
\delta_{\mu} \edf \partial_{\mu} - G^{\nu}_{.
\mu}\frac{\partial}{\partial y^{\nu}} \eqno{(29)}
$$

{\bf{Cartan Linear Connection}}  \\
This connection is defined as
$$
^{\ast}\Gamma^{\alpha}_{. \beta \gamma} \edf \frac{1}{2} g^{\sigma
\alpha} (\delta_{\gamma} g_{\beta \sigma}+ \delta_{\beta} g_{ \sigma
\gamma}
 -\delta_{\sigma} g_{\beta \gamma}) \eqno{(30)}
$$
which can be shown to transform according to the linear connection
transformation,
$$
^{\ast}\bar\Gamma^{\alpha}_{. \beta \gamma}= \frac{\partial \bar
x^{\alpha}}{\partial x^{\mu}}\frac{\partial x^{\nu}}{\partial \bar
x^{\beta}} \frac{\partial x^\sigma}{\partial \bar x^{\gamma}} ~
^{\ast}\Gamma^{\mu}_{. \nu \sigma} +\frac{\partial \bar
x^{\alpha}}{\partial
x^{\sigma}}\frac{\partial^{2}x^{\sigma}}{\partial \bar x^{\beta}
\partial \bar x^{\gamma}} \eqno{(31)}
$$
This connection is a metric one.
\\

{\bf{Berwald Linear Connection}}\\
 This linear connection is defined by,
$$
^{\ast}G^{\alpha}_{. \beta \gamma} \edf \frac{G^{\alpha}_{.
\gamma}}{\partial y^{\beta}} = G^{\alpha}_{ ~ \gamma:\beta}
=G^{\alpha}_{ ~ : ~ \beta\gamma}, \eqno{(32)}
$$
and can be shown to transform according to (31). This connection is
non-metric. It has no torsion since it is symmetric w.r.t. its lower
two indices.
\section{AP-Space With Finslerian Properties}
{\bf{definition}:} An Absolute Parallelism space with  Finslerian
properties $(M,L_{i})$ is an $n$-dimensional differentiable manifold
$M$ equipped with a set of n-Lagrangian functions $L_{i}=
L_{i}(x,y)$ having the
following properties: \\ \\
1.  $L_{i}(x,y)$ is $C^{\infty}$ on $\tau M (= TM \backslash \{0 \})$.  \\
2. $L_{i}(x,y)> 0, y~ \in~ \tau M , ~ ~ y =\dot{x} $.
\\ 3. $L_{i}(x,y)$ is positively homogenous of degree one, i.e. $P-h(1)$.  \\
4. The vectors defined by
$$
\h{i}_{\mu} (x,y) \edf  \frac{\partial L_{i}}{\partial y^{\mu}}
\eqno{(33)}
$$
are assumed to be linearly independent. We are going to abbreviate
this space by FAP-space. The set of Lagrangian functions $L_{i}
(i=1,2,3...,n)$ is called the fundamental set. The set of functions
$\h{i}_{\mu}(x,y)$ are $P-h(0)$ and are the building blocks of the
FAP-space. It can be easily shown that $\h{i}_{\mu}$ transforms as
components of covariant vectors under the group of general
coordinate transformations.

Now, using Euler's theorem and definition (33), we get
$$
L_{i} = \h{i}_{\mu} y^{\mu}. \eqno{(34)}
$$
Let us define the functions,
 $$ \C{i}_ {\mu \alpha} \edf \frac{\partial
\h{i}_{\mu}}{\partial y^{\alpha}} = L_{i: ~ \mu \alpha}.
\eqno{(35)}$$ Again using Euler's theorem, we find that
$$
\C{i}_ {\mu \alpha} y^\mu = \C{i}_ {\mu \alpha} y^\alpha = 0
\eqno{(36)}
$$

Now, the object $\C{i}_{\mu \alpha} (\edf \frac{\partial
\h{i}_{\mu}}{\partial y^{\alpha}}= \h{i}_{\mu : \alpha } )$ has the
following properties: \\ \\
1- $\C{i}_{\mu \alpha}$ is a tensor of type $(0,2)$. \\
2- $\C{i}_{\mu \alpha}$  is $P-h(-1)$. \\
3- $\C{i}_{\mu \alpha}$ is symmetric with respect to its two tensor
indices, as clear from (35).
\\ \\
{\bf {{Theorem I}:}} "A necessary and sufficient condition for an
FAP-space to be an AP-space is that $\C{i}_{\alpha \beta}$ vanishes
identically". \\
This can be easily proved.  As stated above, $\h{i}_{\alpha}$ are
totally independent, then the matrix $(\h{i}_{\alpha})$ is
non-degenerate. Consequently, we can define $\h{i}^{\mu}$ such that:
$$
\h{i}^{\mu}\h{i}_{\nu} = \delta^{\mu}_{\nu}, ~~ \eqno{(37)},
$$
$$
\h{i}^{\mu}\h{j}_{\mu} = \delta_{ij}, ~~ \eqno{(38)}.
$$

{\bf{{Definitions}:}} let
$$
\C{i}^{\mu}_{. \nu} \edf \frac{\partial \h{i}^{\mu}}{\partial
y^{\nu}} = \h{i}^{\mu}_{ ~ : ~ \nu} \eqno{(39)}
$$
$$
\hat{C}^{\mu}_{. \beta \gamma} \edf \h{i}^{\mu} \C{i}_{\beta
\gamma}, \eqno{(40)}
$$
where $\C{i}_{\beta \gamma}$ is the tensor given by (35).
Multiplying both sides of (40) by $\h{j}_\mu$, then we get the
following relation between (35) and (40),
$$
\h {j}_{\mu} \hat{C}^{\mu}_{. \beta \gamma} = \C{j}_{\beta \gamma}.
\eqno{(41)}
$$
These tensors will be used later. \\

{\bf{A Non-Linear Connection}} \\
Using the transformation of the operator $\frac{\partial}{\partial
\bar{x}}$ [12], as a coordinate vector fields on TM, we can write
$$
\frac{\partial \bar{\h{i}_{\mu}}}{\partial \bar{x}^{\alpha}} =
\frac{\partial x^{\beta}}{\partial{\bar{x}^{\alpha}}} \frac{\partial
\bar{\h{i}_{\mu}}} {\partial {x}^{\beta}} +
\frac{\partial^{2}x^{\beta}}{\partial \bar{x}^{\alpha} \partial
\bar{x}^{\sigma}} \bar{y}^{\sigma}\frac{\partial
\bar{\h{i}_{\mu}}}{\partial y^{\beta}}
$$
$$
= \frac{\partial x^{\beta}}{\partial \bar{x}^{\alpha}}
\frac{\partial}{\partial {x}^{\beta}}(\frac{\partial
x^{\gamma}}{\partial \bar{x}^{\mu}} \h{i}_\gamma) +
\frac{\partial^{2}{x}^{\beta}}{\partial \bar{x}^{\alpha}\partial
\bar{x}^{\sigma}} \bar{y}^{\sigma} \frac{\partial}{\partial
y^\beta}(\frac{\partial x^{\gamma}}{\partial
\bar{x}^{\mu}}\h{i}_{\gamma})
$$
$$
=\frac{\partial^{2}{x}^{\gamma}}{\partial \bar{x}^{\alpha}\partial
\bar{x}^{\mu}}\h{i}_{\gamma}+ \frac{\partial
x^{\beta}}{\partial{\bar{x}^{\alpha}}}\frac{\partial
x^{\gamma}}{\partial{\bar{x}^{\mu}}}\frac{\partial
\h{i}_{\gamma}}{\partial x^{\beta}} +
\frac{\partial^{2}{x}^{\beta}}{\partial \bar{x}^{\alpha}\partial
\bar{x}^{\sigma}} \bar{y}^{\sigma} \frac{\partial
x^{\gamma}}{\partial \bar{x}^{\mu}} \h{i}_{\gamma : \beta},
$$
$$
\frac{\partial \bar{\h{i}_{\mu}}}{\partial \bar{x}^{\alpha}} =
\frac{\partial x^{\gamma}}{\partial \bar{x}^{\mu}}\frac{\partial
x^{\beta}}{\partial \bar{x}^{\alpha}} \h{i}_{\gamma, \beta} +
\frac{\partial^{2}{x}^{\gamma}}{\partial \bar{x}^{\mu}\partial
\bar{x}^{\alpha}}\h{i}_{\gamma}+
\frac{\partial^{2}{x}^{\beta}}{\partial \bar{x}^{\sigma}\partial
\bar{x}^{\alpha}}\bar{y}^{\sigma}\frac{\partial x^{\gamma}}{\partial
\bar{x}^{\mu}}\h{i}_{\gamma : \beta}. \eqno{(42)}
$$
Multiplying both sides by $\bar{\h{i}}^{\nu} \bar{y}^\mu$ we get
after some reductions,
$$
(\bar{y}^{\mu} \bar{\h{i}}^{\nu} ~ \frac{\partial
\bar{\h{i}_{\mu}}}{\partial \bar{x}^{\alpha}}) = \frac{\partial \bar
x^{\nu}}{\partial x^{\epsilon}}\frac{\partial x^{\beta}}{\partial
\bar{x}^{\alpha}}(y^{\gamma} \h{i}^{\epsilon} ~ \frac{\partial
\h{i}_{\gamma}}{\partial x^{\beta}}) +
\frac{\partial^{2}{x}^{\gamma}}{\partial \bar{x}^{\mu}\partial
\bar{x}^{\alpha}}\frac{\partial \bar{x}^{\nu}}{\partial x^\gamma}
\bar{y}^{\mu} + \frac{\partial^{2}{x}^{\beta}}{\partial
\bar{x}^{\alpha}\partial \bar{x}^{\sigma}}\frac{\partial
\bar{x}^{\nu}}{\partial x^\epsilon}\bar{y}^{\sigma} y^{\gamma}
\hat{C}^{\epsilon}_{.\gamma \beta}. \eqno{(43)}
$$
Now, using definition (40) and the properties of $\C{i}_{\gamma
\beta}$, given above, we have,
$$
y^{\gamma} \hat{C}^{\epsilon}_{\gamma \beta} = y^{\gamma}
\h{i}^{\epsilon} \C{i}_{\gamma \beta} =0.
$$
This will cause the vanishing of the last term of (43). Then the
quantities between brackets in this equation  transform as
$$
\bar{N}^{\nu}_{. \alpha} = \frac{\partial \bar{x}^{\nu}}{\partial
x^{\epsilon}}\frac{\partial x^{\beta}}{\partial \bar{x}^{\alpha}}
N^{\epsilon}_{. \beta} + \frac{\partial \bar{x}^{\nu}}{\partial
x^{\gamma}}\frac{\partial^{2} x^{\gamma}}{\partial \bar{x}^{\alpha}
\partial \bar{x}^{\mu}} \bar{y}^{\mu}. \eqno{(44)}
$$
where,
$$
N^{\nu}_{. \alpha} \edf y^{\mu} \h{i}^{\nu} \h{i}_{\mu, \alpha}.
\eqno{(45)}
$$
Comparing (44) with (28), one can conclude that $N^{\nu}_{.
\alpha}$, defined by (45), represents the components of a non-linear
connection
defined in FAP-space.\\ \\

{\bf{ Berwald-Like Linear Connection }} \\
Differentiating both sides of (44) w.r.t. $\bar{y}^{\sigma}$, we get
(note that $\frac{\partial}{\partial y}(\frac{\partial x}{\partial
\bar{x}})=0 $)
$$
\bar{N}^{\nu}_{. \alpha : \sigma} =\frac{\partial
\bar{x}^{\nu}}{\partial x^{\epsilon}}\frac{\partial
x^{\beta}}{\partial \bar{x}^{\alpha}}\frac{\partial
x^{\gamma}}{\partial \bar{x}^{\sigma}}N^{\epsilon}_{. \beta : \gamma
} + \frac{\partial \bar{x}^{\nu}}{\partial
x^{\gamma}}\frac{\partial^{2} x^{\gamma}}{\partial \bar{x}^{\alpha}
\partial \bar{x}^{\mu}} \delta^{\mu}_{\sigma}
$$
$$
=\frac{\partial \bar{x}^{\nu}}{\partial x^{\epsilon}}\frac{\partial
x^{\beta}}{\partial \bar{x}^{\alpha}}\frac{\partial
x^{\gamma}}{\partial \bar{x}^{\sigma}}N^{\epsilon}_{. \beta : \gamma
} + \frac{\partial \bar{x}^{\nu}}{\partial
x^{\gamma}}\frac{\partial^{2} x^{\gamma}}{\partial \bar{x}^{\alpha}
\partial \bar{x}^{\sigma}}. \eqno{(46)}
$$
Comparing (46) with (31), one can conclude that the quantity
$\bar{N}^{\nu}_{. \alpha : \sigma}$ transforms as a linear
connection and will be denoted by
$$
B^{\nu}_{. \alpha \sigma} \edf N^{\nu}_{\alpha : \sigma} =
\frac{\partial}{\partial y^{\sigma}} (y^{\mu} \h{i}^{\nu}\h{i}_{\mu,
\alpha}). \eqno{(47)}
$$
It is clear that $B^{\nu}_{. \alpha \mu}$ is non-symmetric in its
lower two indices. So, it has a torsion. \\ \\

{\bf {Cartan-Like Linear Connection }} \\
The non-linear connection defined by (45) can be used to define the
operator $\delta$, similar to (29),
$$
\delta_{\mu} \edf \partial_{\mu} - N^{\alpha}_{.\mu}
\frac{\partial}{\partial y^{\alpha}}.   \eqno{(48)}
$$
So, if $A_{\alpha}$ is an arbitrary vector, then we can define the
derivative,
$$
\delta_\beta A_\alpha = A_{\alpha ~ \hat{,} ~ \beta} \edf A_{\alpha
, \beta } -
N^{\gamma}_{.\beta} A_{\alpha : \gamma}. \eqno{(49)} $$ \\
Using the operator (48), we can define the object,
$$
\Gamma^{\mu}_{. \alpha \beta} \edf {\h{i}}^{\mu} \h{i}_{\alpha
\hat{,} \beta}. \eqno{(50)}
$$
Such quantities can be shown to transform according to (31). So, the
set of quantities, given by (50) represents the components of a
linear connection, different from (47), and of Cartan type. It is
non-symmetric w.r.t. its lower two indices. Consequently, it admits
a torsion.  \\

{\bf{Covariant V- and H- Derivatives}:} \\
Let us define the following vertical (V-) covariant derivative
$$
A_{\mu}|_{\nu} \edf A_{\mu : \nu} + A_{\alpha}\hat{C}^{\alpha}_{.\mu
\nu}, \eqno{(51)}
$$
where,
$$
A_{\mu : \nu} = \frac{\partial A_{\mu}}{\partial y ^{\nu}}
\eqno{(52)}
$$
and
$$
\hat{C}^{\alpha}_{.\mu \nu} = {\h{i}}^{\alpha} \frac{\partial^{2}
L_{i}}{\partial y^{\mu}
\partial y^{\nu}} \eqno{(53)}
$$
Now, for the building blocks of the FAP-space, using (51) we get
$$
\h{i}_{\alpha}|_{\beta}  \equiv 0. \eqno{(54)}
$$
This implies that $\h{i}_{\alpha}$ are parallel displaced along a
certain path characterized by $\hat{C}^{\gamma}_{. \alpha \beta}$
(absolute parallelism). Consequently, using (37), we can show that,
$$
\h{i}^{\mu}|_{\nu} \equiv 0. \eqno{(55)}
$$

We can also define the H-covariant derivative as
$$
A_{\alpha \| \beta} = A_{\alpha \hat{,} \beta} - A_{\mu}
\Gamma^{\mu}_{. \alpha \beta}, \eqno{(56)}
$$
where $A_{\alpha \hat{,} \beta}$ is given by (49) and
$\Gamma^{\mu}_{. \alpha \beta}$ is the linear connection defined by
(50). Consequently, we get
$$
\h{i}_{\alpha \| \beta} \equiv 0. \eqno{(57)}
$$
Similarly, we can write
$$
\h{i}^{\beta}_{~ \| \gamma} \equiv 0. \eqno{(58)}
$$
The relation (57) shows that $\h{i}_\alpha$ are parallel displaced
along a certain path characterized by the linear connection (50). \\
\\ \\

{\bf Introduction of a Metric}\\
So far, all the above calculations have been performed without using
a metric. This is achieved by using the building blocks of the
FAP-space. But due to the importance of the metric tensor in
physical applications and in order to extend the geometric
structure, it is preferable to introduce a metric in this stage. For
this reason, consider the following quantities: Let us define the
second order symmetric tensors
$$
g_{\mu \nu} \edf  \h{i}_{\mu} \h{i}_{\nu} ,  \eqno{(59)}
$$
$$
g^{\alpha \beta}  \edf  \h{i}^{\alpha} \h{i}^{\beta}.  \eqno{(60)}
$$
Consequently, using the relations (37) and (38) on the building
blocks of the FAP-space, we can write
$$
g^{\alpha \mu}g_{\beta \mu} = \delta^{\alpha}_{\beta}.
$$
Consider now the quantity,
$$
C_{\alpha \beta \gamma} \edf \frac{1}{2} \frac{\partial g _{\beta
\gamma}}{\partial y^\alpha}
$$
$$
{\ }{\ }{\ }= \frac{1}{2} \frac{\partial}{\partial
y^{\alpha}}(\h{i}_{\beta} \h{i}_{\gamma})
$$
$$
{\ }{\ }{\ }= \frac{1}{2} (\h{i}_{\beta} \h{i}_{\gamma : \alpha}+
\h{i}_{\beta : \alpha} \h{i}_{\gamma})
$$
$$
{\ }{\ }{\ }= \frac{1}{2} (\C{i}_{\gamma \alpha} \h{i}_{ \beta}+
\C{i}_{\beta  \alpha} \h{i}_{\gamma}). \eqno{(61)}
$$
 If we use the definition,
 $$
 \hat{C}_{\alpha \beta \gamma} \edf \C{i}_{\alpha  \beta}
 \h{i}_{\gamma}
 $$
 then we can write,
$$
\hat{C}_{\alpha \beta \gamma} = \hat{C}_{\alpha (\beta \gamma)}
+\hat{C}_{\alpha [\beta \gamma]}, \eqno{(62)}
$$
where,
$$
\hat{C}_{\alpha (\beta \gamma)} \edf \frac{1}{2}(\hat{C}_{\alpha
\beta \gamma}+ \hat{C}_{\alpha \gamma \beta} ), \eqno{(63)}
$$
$$
\hat{C}_{\alpha [\beta \gamma]} \edf \frac{1}{2}(\hat{C}_{\alpha
\beta \gamma}- \hat{C}_{\alpha \gamma \beta} ). \eqno{(64)}
$$
From (61)and (63)  we can write $C_{\alpha \beta \gamma} =
\hat{C}_{\alpha (\beta \gamma)}$. The tensor
$C_{\alpha \beta \gamma}$ has the following properties: \\
1- It is a tensor of type (0,3), symmetric w.r.t. all indices. \\
2- It is $P-h(-1)$. \\
3- $C_{\alpha \beta \gamma} y^{\alpha}= C_{\alpha \beta \gamma}
y^{\beta}= C_{\alpha \beta \gamma} y^{\gamma} = 0 $ (using Euler's
theorem ). It is to be considered that the tensor (61) has the same
properties of the tensor given by (22), although it is defined in
terms of the building blocks of the FAP-space. For this reason the
two tensors are referred to using
the same symbol. \\ \\

Let us define the energy of FAP-space,
 $$ E \edf
\frac{1}{2}L^{2} = \frac{1}{2} \sum_{i} L_{i}L_{i}=
\frac{1}{2}\h{i}_{\alpha}\h{i}_{\beta} y^{\alpha}y^{\beta},
\eqno{(65)}
$$
such that $E$ is $P-h(2)$, then
$$
\frac{\partial E}{\partial y^{\sigma}} = \frac{1}{2} \h{i}_{\alpha}
\h{i}_{\beta} \delta^{\alpha}_{\sigma} y^{\beta} +
\frac{1}{2}\h{i}_{\alpha} \h{i}_{\beta} \delta^{\beta}_{\sigma}
y^{\alpha} + \frac{1}{2}(\h{i}_{\alpha} \h{i}_{\beta :
\sigma}+\h{i}_{\beta}\h{i}_{\alpha : \sigma}) y^{\alpha} y^{\beta}
$$
$$
= \h{i}_{\sigma}\h{i}_{\beta} y^{\beta}+ C_{\alpha \beta
\sigma}y^{\alpha}y^{\beta}
$$
$$
= \h{i}_{\sigma}\h{i}_{\beta} y^{\beta} ~ ~ ~ ~ {\it (using (23))}.
$$
$$
\frac{\partial^{2}E }{\partial y^{\sigma} y^{\epsilon}} =
\h{i}_{\sigma : \epsilon}  \h{i}_{\beta}  y^{\beta} + \h{i}_{\beta :
\epsilon}  \h{i}_{\sigma}  y^{\beta} + \h{i}_{\sigma}
\h{i}_{\epsilon}
$$
$$
=\h{i}_{\sigma} \h{i}_{\epsilon} + 2 C_{\epsilon \beta
\sigma}y^{\beta}
$$
$$
=g_{\sigma \epsilon}.  ~ ~ ~ ~ {\it (using (23))}. \eqno {(66)}
$$
Now, the object $g_{\mu \nu}$ has the following properties: \\
1- It is a symmetric tensor of type (0,2). \\
2- It is non-degenerate since $\h{i}$ is non-degenerate. \\
3- It is $P-h(0)$, positively homogenous of degree 0 in y. \\
4- It is derivable from the energy of the FAP-space,
$E=\frac{1}{2}\sum_{i}L_{i}L{i}$, as given by (66).\\
So, it can be used as a metric tensor of a space associated with the
FAP-space.\\ \\

{\bf Metricity Conditions}\\
 Equation (54) implies that
$$
g_{\alpha \beta}|_{\gamma} \equiv 0. \eqno{(67)}
$$
But we have
$$
g^{\mu \beta}g_{\mu \alpha} = \delta^{\beta}_{\alpha}
$$
Consequently, we get
$$
g^{\epsilon \beta}|_{\sigma} \equiv 0 \eqno{(68)}
$$
From (67) and (68) a metricity condition is automatically
satisfied. This implies that the operator of raising and lowering
indices commutes with the V-covariant differential operator. In
other words, one can usee $g^{\mu \nu}$ and $g_{\alpha \beta}$ for
raising and lowering tensor indices, under V-covariant operator
sign, respectively.

Also, as a direct consequence of (58) and (59) we can write
$$
g_{\mu \nu \| \sigma} \equiv 0 , \eqno{(69)}
$$
$$
g^{\mu \nu}_ { ~ ~ \| \sigma} \equiv 0 , \eqno{(70)}
$$
which shows that (50) defines a  metric linear connection. This
shows that the tensor $g_{\alpha \beta}$ and its conjugate can be
used to lower and raise tensor indices under the H-covariant
differentiation sign, respectively.
\\ \\
{\bf A Missing Condition:} Although we have shown that the metric
can be derived from the energy of the FAP-space, as given by (66),
it implies a certain condition as shown below.\\ {\bf{Theorem II}:}
In case of $\hat{C}_{\alpha [\beta \gamma]=0}$ the FAP-space has an
associated Finsler space whose metric is given by
$$
g_{\alpha \beta} = \frac{1}{2} \frac{\partial^{2}L^{2}}{\partial
y^{\alpha} \partial y^{\beta}}
$$
{\bf proof}
$$
g_{\mu \nu} = \frac{1}{2}\frac{\partial^{2}L^{2} }{\partial y^{\mu}
y^{\nu}} = \frac{\partial}{\partial y^{\mu}}(\frac{\partial.
L_{i}}{\partial y^{\nu}} L_{i}),
$$
$$
= \frac{\partial L_{i}}{\partial y^{\nu}}\frac{\partial
L_{i}}{\partial y^{\mu}} + L_{i}\frac{\partial^{2} L_{i}}{\partial
y^{\mu}y^{\nu}}
$$
$$
= \h{i}_{\mu} \h{i}_{\nu} + \h{i}_{\alpha}y^{\alpha} \C{i}_{\mu \nu}
$$
$$
= \h{i}_{\mu} \h{i}_{\nu} + y^{\alpha}\hat{C}_{\mu \nu \alpha}
$$
$$
= \h{i}_{\mu} \h{i}_{\nu} + y^{\alpha}(\hat{C}_{\mu (\nu \alpha)
)}+\hat{C}_{\mu [\nu \alpha]})
$$
$$
= \h{i}_{\mu} \h{i}_{\nu} + y^{\alpha} {C}_{\mu \nu \alpha} +
y^{\alpha}\hat{C}_{\mu [\nu \alpha]}
$$
$$
= \h{i}_{\mu} \h{i}_{\nu}. \eqno{(71)}
$$

Therefore the above theorem is proved.\\ \\
\section{Concluding Remarks}
1- In the present article, the bases of a parallelisable structure
with Finslerian properties, have been suggested. It possesses both
advantages of the conventional AP-geometry and Finser geometry.\\
2- Theorem I gives the necessary and sufficient condition for an
FAP-structure to be an AP-one. The vanishing of the tensor (35) is a
mathematical expression for this condition. This condition would be
very useful in applications. For a physical application written in
the FAP-space, it is easy to get its AP-picture ,and consequently
its Riemannian one, just by using this condition. This would help in
attributing some physical
meaning to Finslerian geometric objects. \\
3- All geometric quantities, of the FAP-structure given in the
present work, are defined in terms of the building blocks of this
structure. It is worth of mention that the linear connection given
by (50) is similar to that defined in reference [14], but in the
present work we use the non-linear connection (45) that is defined
in terms of the building blocks of the FAP-structure.\\
4- The condition for the metric tensor (59) to be a Finslerian
metric is given by theorem II. This facilitates comparison between
any physical application written in the FAP-structure and its
Finsierian picture, if any.\\
5- In the structure developed, we meant by the torsion the
anti-symmetric part of any linear connection.\\
6- In the present part of the work, we focused on the building
blocks, connections, covariant derivatives and metric. The study of
curvatures, the W-tensor [13], torsion and identities will be given
in the next
part of this work.\\
\section*{Acknowledgments}
The author would like to thank Professor N.L. Youssef, Dr. S. Abed
and Mr. A. Sid-Ahmed for many discussions and comments.

\section*{References}
{[1] Y. Sofue and V. Rubin (2001) Ann.Rev.Astron.Astrophys. {\bf 39}, 137-174; astro-ph/0010594.}\\
 {[2] P.D. Mannheim (2006) Prog.Part.Nucl.Phys. {\bf 56},
340-445; astro-ph/0505266.}\\
 {[3] J.L. Tonry, P.B. Schmidit et al.
(2003) Astrophys.J. {\bf 594}, 1-24; astro-ph/0305008.}\\
 {[4] S.Weinberg (1972) {\it "Gravitation and Cosmology"}, John Wiley \&
Sons.}\\
 {[5] A. Einstein (1955) {\it "The Meaning of Relativity"},
App. II, Princeton, 5th ed.}\\
 {[6] G. Shipov (1998) {\it "A Theory
of Physical Vacuum"}, Moscow.}\\
 {[7] F.I. Mikhail and M.I. Wanas
(1977) Proc.Roy.Soc.Lond. {\bf A356}, 471-481.}\\
 {[8] R. Miron and
M. Anastasiei (1997) {\it "Vector Bundles and Lagrange Spaces

with Applications to Relativity"}, Balakan Press.}\\
 {[9] G.S. Assanov
and S.F. Ponomarenko (1989) {\it "Finsler Bundles Over Space-Time:

Associated Gauge Fields and Connections"}, Nauka.}\\
 {[10] M.I. Wanas (2000) Turk.J.Phys. {\bf 24}, 473-488; gr-qc/0010099.}\\
 {[11] M.I. Wanas (2001) Cercet.Stiin.Ser.Mat. {\bf 10}, 297-309;
gr-qc/0209050.}\\
 {[12] D. Bao, S.S. Chern and Z. Shen (2000) {\it
"An Introduction to Riemann-Finsler

Geometry"}, Springer.}\\
{[13] N.L. Youssef and A.M. Sid-Ahmed (2006) Rep.Math.Phys. {\bf 60}, 39-53; gr-qc/0604111.}\\
{[14] M.I. Wanas, N.L. Youssef and A.M. Sid-Ahmed (2007) arxiv 0704.2001.}\\

\end{document}